\documentclass{PoS2}

\usepackage{amsmath, amsthm, amssymb} 
\usepackage{cite}

\newcommand*\pFqskip{8mu}
\catcode`,\active
\newcommand*\pFq{\begingroup
        \catcode`\,\active
        \def ,{\mskip\pFqskip\relax}%
        \dopFq
}
\catcode`\,12
\def\dopFq#1#2#3#4#5{%
        {}_{#1}F_{#2}\biggl[\genfrac..{0pt}{}{#3}{#4};#5\biggr]%
        \endgroup
}
\title{{\footnotesize{DESY 18-117, DO-TH 18/15}}\\
The $\rho$ parameter at three loops and elliptic integrals}

\ShortTitle{The $\rho$ parameter at three loops and elliptic integrals}

\author{J. Bl\"umlein$^a$, \speaker{A. De Freitas}$^a$, M.~van~Hoeij$^b$, E.~Imamoglu$^b$, 
P.~Marquard$^a$, and~C.~Schneider$^c$ \\
\llap{$^a$}Deutsches Elektronen--Synchrotron, DESY, \\
                   Platanenallee 6, D-15738 Zeuthen, Germany. \\
        \llap{$^b$}Department of Mathematics, Florida State University, \\
                   208 Love Building, 1017 Academic Way, Tallahassee, FL 32306-4510, USA.\\
        \llap{$^c$}Research Institute for Symbolic Computation (RISC), Johannes Kepler 
        University, \\ 
                   Altenbergerstra\ss{}e 69, A--4040, Linz, Austria.} 

\abstract{We describe the analytic calculation of the master integrals required to compute 
the two-mass three-loop corrections to the $\rho$ parameter. In particular, we present 
the calculation of the master integrals for which the corresponding differential equations 
do not factorize to first order. The homogeneous solutions to these differential 
equations are obtained in terms of hypergeometric functions at rational argument. These 
hypergeometric functions can further be mapped to complete elliptic integrals, and the 
inhomogeneous solutions are expressed in terms of a new class of integrals of  combined 
iterative non-iterative nature.}

\FullConference{ Loops and Legs in Quantum Field Theory - LL 2018,\\
		 29 April - 4 May 2018 \\
		 St. Goar, Germany }

\begin{document}

\section{Introduction}

\noindent
The calculation of multi-loop Feynman integrals constitutes a crucial step required for the 
computation of quantum corrections to different standard model processes occurring at the LHC 
and other collider experiments. 
Considerable progress has been made in this regard in the past few decades, and many 
Feynman integrals have been computed using a variety of methods.\footnote{For a recent 
review on available calculation methods see Ref.~\cite{Blumlein:2018cms}.} 
In particular, integration by parts identities (IBP) have been used to express all required Feynman integrals in terms of a small set of so called master integrals \cite{IBP}. 
The master integrals can then be calculated by taking their derivatives with respect to the invariants of the problem, which leads to an expression that can be rewritten in terms of the master integrals themselves by inserting the IBPs. This leads to a system of differential equations for the master integrals, which can be decoupled and, given appropriate boundary conditions, one may then try to solve \cite{DEQ}.
In many cases, the decoupled equations turn out to be first order factorizable, and the master integrals can be expressed in terms of iterated integrals, such as the 
harmonic polylogarithms \cite{Remiddi:1999ew}, Kummer-Poincar\'e iterated 
integrals \cite{Moch:2001zr,Ablinger:2013cf},
cyclotomic polylogarithms \cite{Ablinger:2011te}, and iterated integrals with 
squared-root-valued letters in the 
alphabet \cite{Ablinger:2014bra}, among others \cite{TWOMASS}. Also associated nested 
sum representations are obtained \cite{HSUM,Ablinger:2013cf,Ablinger:2014bra,TWOMASS}
and special constants appear in these
representations, see e.g. \cite{Blumlein:2009cf}. 

There are many physical 
problems that have been solved entirely 
in terms of these types of functions, particularly, problems involving only massless particles or planar Feynman diagrams. 
On the other hand, when trying to solve problems involving massive particles and/or 
non-planar diagrams, one may encounter Feynman integrals for which the corresponding differential equation turns out not to 
be first order factorizable, 
cf.~\cite{BLOCH2,Adams:2015gva,Adams:2015ydq,Adams:2014vja,Adams:2016xah,Ablinger:2017bjx,
MullerStach:2011ru,
Adams:2013nia,
Remiddi:2013joa,
Bloch:2014qca,
Sogaard:2014jla,
Bloch:2016izu,
Remiddi:2016gno,
Bonciani:2016qxi,
Passarino:2016zcd,
vonManteuffel:2017hms,
Primo:2017ipr,
Adams:2017ejb,
Bogner:2017vim,
Remiddi:2017har,
Lee:2017qql,
Bourjaily:2017bsb,
Hidding:2017jkk,
Broedel:2017kkb,
Broedel:2017siw,
Adams:2018yfj,
Broedel:2018iwv,
Groote:2018rpb,
Adams:2018bsn,
Adams:2018kez,
Lee:2018ojn,
Broedel:2018rwm,
Adams:2018ulb}.
Feynman integrals obeying differential equations that can be factorized to first order, except for one irreducible term of second order, represent the next level of complexity among the integrals arising in many problems of interest in perturbative calculations. One such problem turns out to be the two-mass three-loop corrections to the $\rho$ parameter. This problem is ideal for the study of the computation of this type of Feynman integrals because it has a particularly nice feature, namely, the master integrals for which the differential equations contain a term of second order are almost the last ones that need to be solved. Other problems, which require the solutions of master integrals such as the sunrise or the kite integrals 
\cite{BLOCH2,Adams:2015gva,Adams:2015ydq,Adams:2014vja,Adams:2016xah}, can be more cumbersome, since these integrals are usually the first ones that need to be solved and their solutions reappear throughout the rest of differential equation system, with new integrations over these solutions at each step. The fact that this does not happen in the case of the 
two-mass three-loop contributions to the $\rho$ parameter means that it is simpler to obtain a fully analytic result for the physical quantity.

\begin{figure}[t]
\begin{minipage}[c]{0.15\linewidth}
     \includegraphics[width=0.9\textwidth]{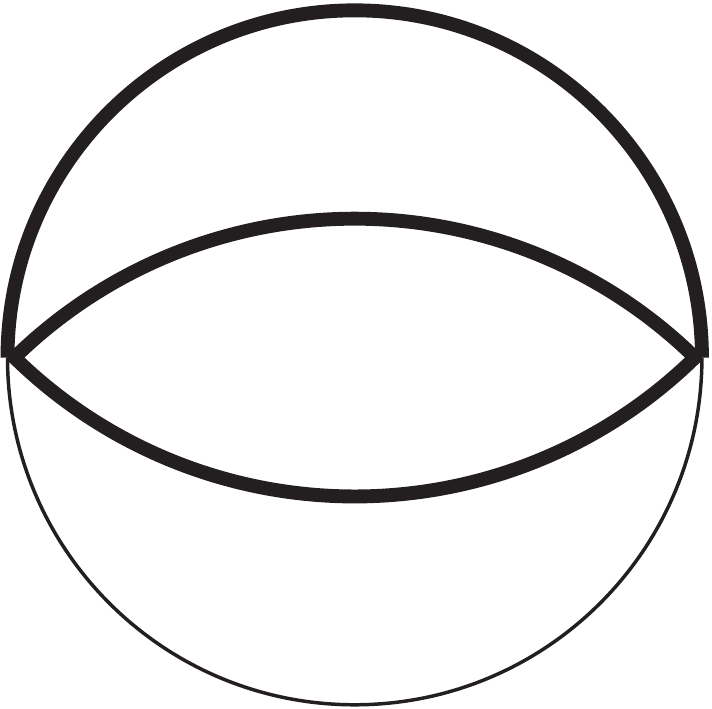}
\vspace*{-3mm}
\begin{center}
{\footnotesize $J^{(3)}_{8a}$}
\end{center}
\end{minipage}
\hspace*{1mm}
\begin{minipage}[c]{0.15\linewidth}
     \includegraphics[width=0.9\textwidth]{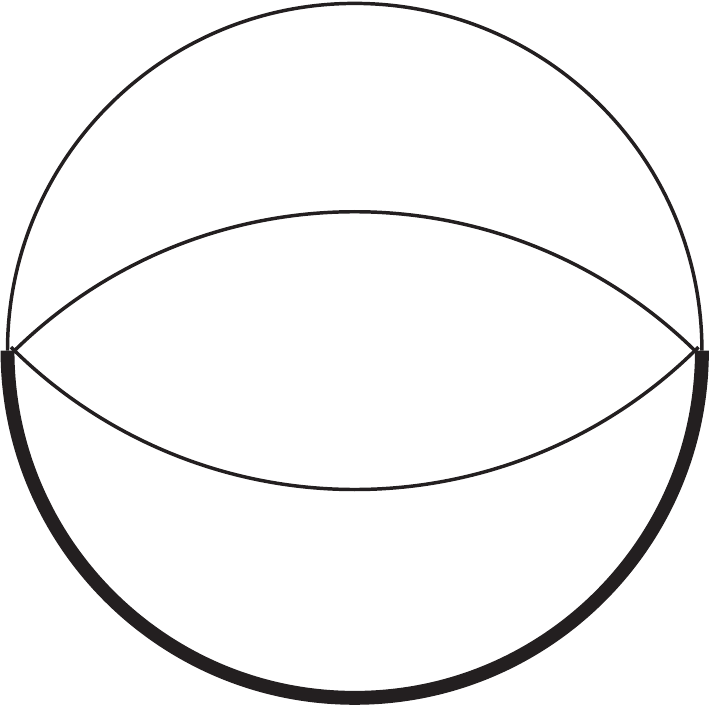}
\vspace*{-3mm}
\begin{center}
{\footnotesize $J^{(3)}_{8b}$}
\end{center}
\end{minipage}
\hspace*{1mm}
\begin{minipage}[c]{0.15\linewidth}
     \includegraphics[width=0.9\textwidth]{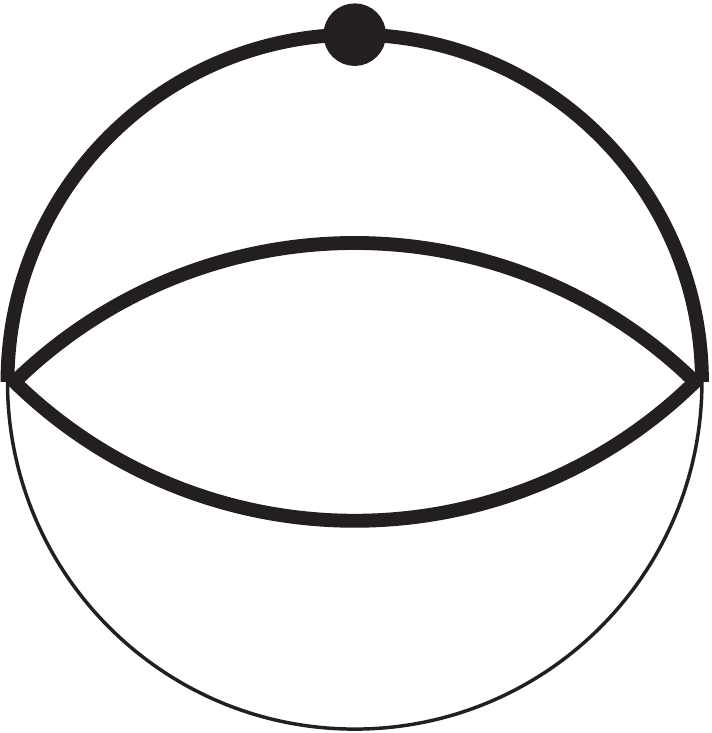}
\vspace*{-3mm}
\begin{center}
{\footnotesize $J^{(3)}_{9a}$}
\end{center}
\end{minipage}
\hspace*{1mm}
\begin{minipage}[c]{0.15\linewidth}
     \includegraphics[width=0.9\textwidth]{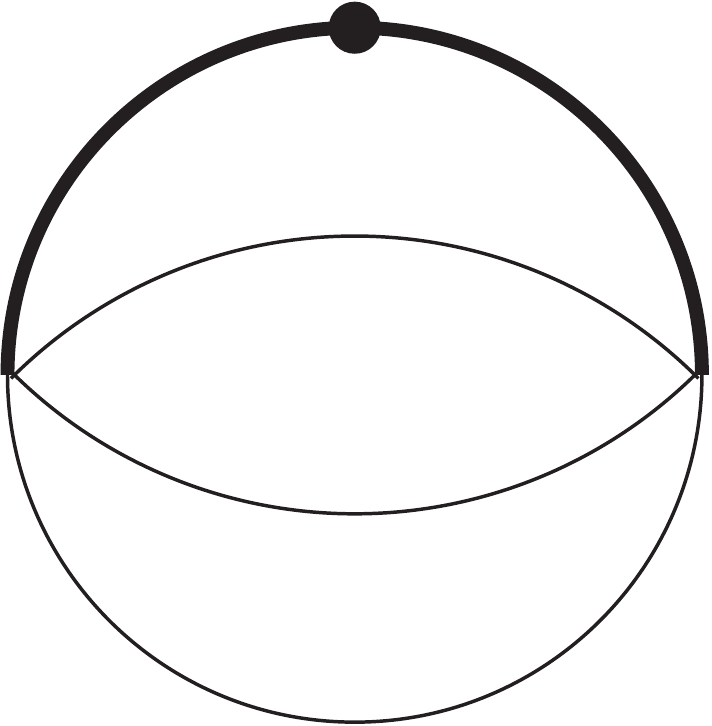}
\vspace*{-3mm}
\begin{center}
{\footnotesize $J^{(3)}_{9b}$}
\end{center}
\end{minipage}
\hspace*{1mm}
\begin{minipage}[c]{0.15\linewidth}
     \includegraphics[width=0.9\textwidth]{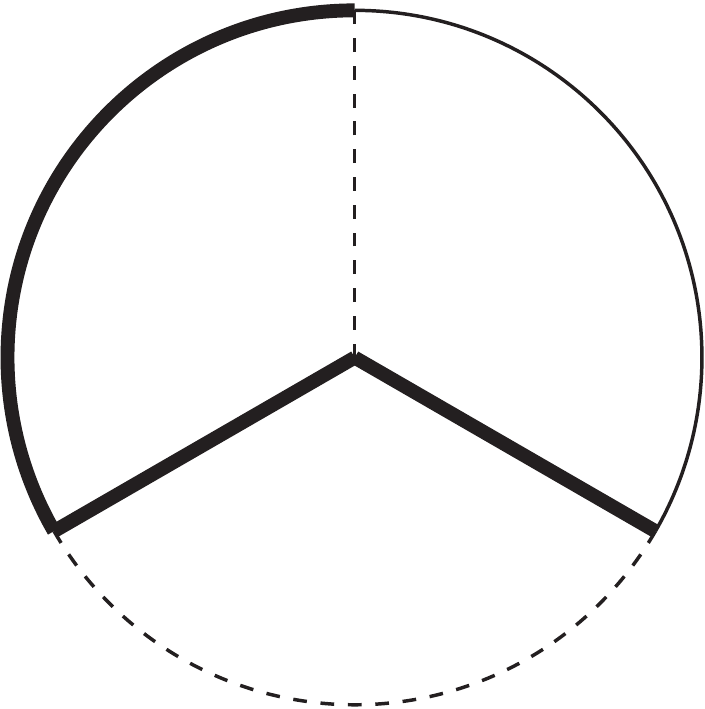}
\vspace*{-3mm}
\begin{center}
{\footnotesize $J^{(3)}_{10a}$}
\end{center}
\end{minipage}
\hspace*{1mm}
\begin{minipage}[c]{0.15\linewidth}
     \includegraphics[width=0.9\textwidth]{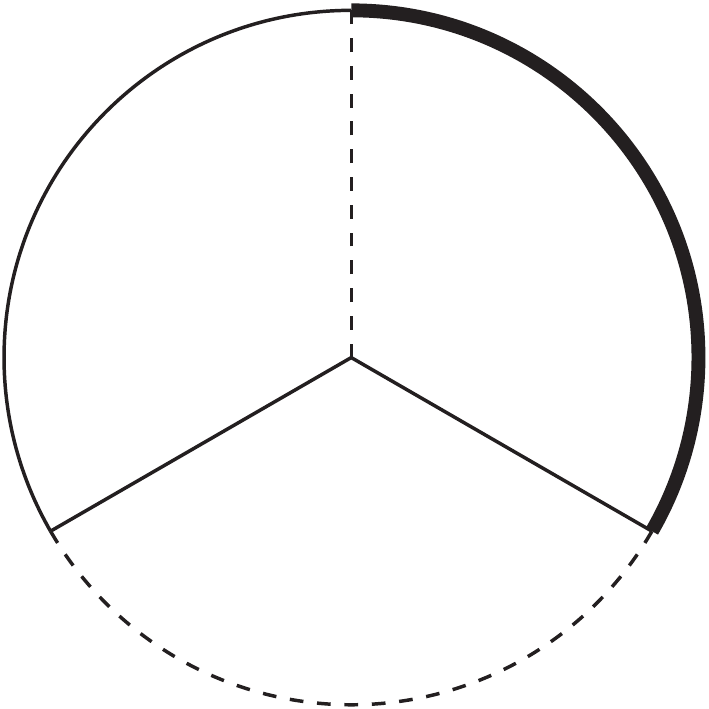}
\vspace*{-3mm}
\begin{center}
{\footnotesize $J^{(3)}_{10b}$}
\end{center}
\end{minipage}
\label{fig:MIs}
\caption{The six master integrals required for the calculation of the two-mass three-loop contributions to the $\rho$ parameter which obey differential equations that are not factorizable to first order. Dashed lines represent massless propagators, while solid lines represent massive propagators (thin lines for the smaller mass and thick lines for the larger one). A dot on a line represents a squared propagator.} 
\end{figure}

The $\rho$ parameter is an important quantity in the standard model \cite{Ross:1975fq}
that measures the 
relative strength between the neutral current and charged current interaction and is given by
\begin{equation}
\rho = \frac{M_W^2}{M_Z^2 \cos^2(\theta_W)}~,
\end{equation}
which at tree level is equal to 1, and receives quantum corrections
\begin{equation}
\rho = 1 + \Delta \rho,
\end{equation}
given by
\begin{equation}
\Delta \rho = \frac{\Sigma_Z(0)}{M_Z^2} - \frac{\Sigma_W(0)}{M_W^2},
\end{equation}
where $\Sigma_Z(0)$ and $\Sigma_W(0)$ are the transverse parts of the $Z$ and $W$ boson propagators, respectively, 
which are defined by
\begin{equation}
\Sigma_{W/Z}(0) = \frac{g_{\mu \nu}}{d} \Pi_{W/Z},
\end{equation}
where $\Pi_{W/Z}$ are the corresponding polarization functions. The quantum loop corrections as an expansion in the strong coupling constant $\alpha_s$ can be written as 
\begin{equation}
\Delta \rho = \frac{3 G_F m_t^2}{8 \pi^2 \sqrt{2}} \biggl(
\delta^{(0)}+\frac{\alpha_s}{\pi} \delta^{(1)}
+\left(\frac{\alpha_s}{\pi}\right)^2 \delta^{(2)}
+{\cal O}(\alpha_s^3)\biggr).
\label{eq:deltarho}
\end{equation}
Some of the most recent computations of these quantum corrections can be found in \cite{Avdeev:1994db,Chetyrkin:1995ix,Boughezal:2004ef,Chetyrkin:2006bj,Boughezal:2006xk,Grigo:2012ji}. In the calculation of the two-mass contribution to the three-loop term $\delta^{(2)}$ presented in \cite{Grigo:2012ji}, it was found that all but six of the master integrals required to obtain this quantity could be computed in terms of harmonic polylogarithms depending on the ratio of the two masses,
\begin{equation}
x = \frac{m_1}{m_2}, \quad \quad m_1 < m_2 \quad \Rightarrow \quad 0<x<1.
\end{equation}
The remaining six master integrals are depicted in Figure~\ref{fig:MIs}, and can be 
expanded 
up to order $\varepsilon^0$ (the calculation is done using dimensional regularization, where the dimension $D$ is given by $D=4-2 \varepsilon$) as follows
\begin{equation}
J_k^{(3)}(x) = \frac{1}{\varepsilon^3} g_{k,-3}(x) + \frac{1}{\varepsilon^2} g_{k,-2}(x) + 
\frac{1}{\varepsilon} g_{k,-1}(x) + f_k(x) + {\cal O}(\varepsilon)\, ,
\end{equation}
where $k$ labels the different master integrals. We use the same notation as the authors of Ref. \cite{Grigo:2012ji}, i.e., in the case of the six master integrals under consideration, $k \in \{8a,8b,9a,9b,10a,10b\}$. It turns out that the pole terms $g_{k,-i}$ ($i=1,2,3$) of these six master integrals can also be expressed entirely in terms of harmonic polylogarithms in the variable $x$, while the constant terms $f_k$ obey a non-factorizable second order differential equation in the case of $k \in \{8a,8b,9a,9b\}$, and in the case of $k \in \{10a,10b\}$ we have a first order differential equation where the previous integrals appear in the inhomogeneities. 

%-----------------------------------------------------------------------------------------------------
%-----------------------------------------------------------------------------------------------------

In Ref. \cite{Grigo:2012ji}, the authors obtained these master integrals in terms of 
series expansions around $x=0$ and $x=1$, which for numerical purposes turns out to be 
enough, since these expansions are very well behaved and overlap over a wide range of $x 
\in (0,1)$. Here we present the calculation of these integrals in analytic form 
\cite{Ablinger:2017bjx}. We will see that the homogeneous part of the differential 
equations can be solved in terms of elliptic integrals, and the inhomogeneous solutions 
can be written in terms of a new type of iterated integrals where specific 
configurations out of the complete elliptic integrals can be interpreted as new letters in 
the contributing alphabet.
%%%%%%%%%%%%%%%%%%%%%%%%%%%%%%%%%%%%%%%%%%%%%%%%%%%%%%%%%%%%%%%%%%%%%%%%%%%%%%%%%%%%%%%%%%%%%%%%%%%%%%%%%%%%%%%%%%%%%%%%%%%%%%%%%%%%%%%%%%%%%%%%%%%
\section{The differential equations}

\noindent
The constant terms $f_{8a}(x)$ and $f_{9a}(x)$ of the master integrals $J^{(3)}_{8a}(x)$ and $J^{(3)}_{9a}(x)$ obey the following system of differential equations
%-----------------------------------------------------------------------------------------------------
\begin{eqnarray}
\renewcommand{\arraystretch}{1.6}
\frac{d}{dx}
\left(
\begin{array}{c}
f_{8a}(x) \\
f_{9a}(x) 
\end{array}
\right)
= \left(
\begin{array}{cc}
\frac{4}{x} & \frac{6}{x} \\
\frac{4(x^2-3)}{x(x^2-9)(x^2-1)} &
\frac{2(x^4-9)}{x(x^2-9)(x^2-1)} 
\end{array}
\right) \otimes
\left(
\begin{array}{c}
f_{8a}(x) \\
f_{9a}(x) 
\end{array}
\right)
+
\left(
\begin{array}{c}
R_{8a}(x) \\
R_{9a}(x) 
\end{array}
\right),
\end{eqnarray}
\renewcommand{\arraystretch}{1.0}
%-----------------------------------------------------------------------------------------------------
where
%-----------------------------------------------------------------------------------------------------
\begin{eqnarray}
R_{8a}(x) &=& -\frac{195}{4 x}-4 x+\frac{x^3}{4}
%-\frac{15 (13 + 16 x^2 - x^4)}{4 x} 
+ 3 x \left(24 - x^2\right) \ln(x) - 18 x \ln^2(x),
\\
R_{9a}(x) &=& \frac{1}{\left(x^2-9\right) \left(x^2-1\right)} \biggl[
\frac{585}{4 x}+\frac{621}{4} x-\frac{1255}{12} x^3+\frac{157}{12} x^5
- \frac{16}{3} x^3 \ln^3(x)
\nonumber\\ &&
+ 2 x \left(45 - 17 x^2 + 2 x^4\right) \ln^2(x)
- x \left(324 - 145 x^2 + 15 x^4\right) \ln(x)
\biggr],
\end{eqnarray}
%-----------------------------------------------------------------------------------------------------
while the constant terms $f_{8b}(x)$ and $f_{9b}(x)$ of the master integrals $J^{(3)}_{8b}(x)$ and $J^{(3)}_{9b}(x)$ satisfy
%-----------------------------------------------------------------------------------------------------
\begin{eqnarray}
\renewcommand{\arraystretch}{1.6}
\frac{d}{dx}
\left(
\begin{array}{c}
f_{8b}(x) \\
f_{9b}(x) 
\end{array}
\right)
= \left(
\begin{array}{cc}
\frac{4}{x} & \frac{2}{x} \\
\frac{4(3x^2-1)}{x(9x^2-1)(x^2-1)} &
\frac{2(9x^4-1)}{x(9x^2-1)(x^2-1)} 
\end{array}
\right) \otimes
\left(
\begin{array}{c}
f_{8b}(x) \\
f_{9b}(x) 
\end{array}
\right)
+
\left(
\begin{array}{c}
R_{8b}(x) \\
R_{9b}(x)
\end{array}
\right),
\end{eqnarray}
\renewcommand{\arraystretch}{1.0}
%-----------------------------------------------------------------------------------------------------
with
%-----------------------------------------------------------------------------------------------------
\begin{eqnarray}
R_{8b}(x) &=& \frac{15}{4 x}-4 x-\frac{13}{4} x^3
%-\frac{15 (-1+16 x^2+13 x^4)}{4 x}
+9 x \left(8+15 x^2\right) \ln(x) - 18 \left(x+6 x^3\right) \ln^2(x),
\\
R_{9b}(x) &=& 
\frac{1}{\left(9x^2-1\right) \left(x^2-1\right)} \biggl[
-\frac{15}{4 x}+\frac{397}{4} x-\frac{925}{4} x^3-\frac{297}{4} x^5
-16 x^3 \left(8-27 x^2\right) \ln^3(x)
\nonumber\\ &&
+6 x \left(5+37 x^2-144 x^4\right) \ln^2(x)
-3 x \left(36-35 x^2-195 x^4\right) \ln(x)
\biggl].
\end{eqnarray}
%-----------------------------------------------------------------------------------------------------
By applying decoupling algorithms \cite{Zuercher:94,ORESYS,NewUncouplingMethod} 
one obtains the following 
scalar differential equation 
%-----------------------------------------------------------------------------------------------------
\begin{eqnarray}
\left(\frac{d^2}{dx^2}
+\frac{9-30 x^2+5 x^4}{x(x^2-1)(9-x^2)} \frac{d}{dx}
+\frac{8 (3-x^2)}{(9-x^2)(x^2-1)}\right) f_{8a}(x) 
&=& 
N_{8a}(x)
\label{eq:one}
\end{eqnarray}
%-----------------------------------------------------------------------------------------------------
with
%-----------------------------------------------------------------------------------------------------
\begin{eqnarray}
N_{8a} &=& 
\frac{1}{\left(9-x^2\right) \left(x^2-1\right)} \biggl[
32 x^2 \ln^3(x)
+12 \left(9-13 x^2-2 x^4\right) \ln^2(x)
\nonumber\\ &&
-6 \left(54-62 x^2-x^4-x^6\right) \ln(x)
+\frac{1161}{2}-\frac{251}{2} x^2-\frac{61}{2} x^4-\frac{9}{2} x^6
\biggr]
\end{eqnarray}
%-----------------------------------------------------------------------------------------------------
together with the equation
%-----------------------------------------------------------------------------------------------------
\begin{eqnarray}
f_{9a}(x) &=& N_{9a}(x) - \frac{2}{3} f_{8a}(x) + \frac{x}{6} \frac{d}{dx} f_{8a}(x),
\label{eq:f9aa}
\end{eqnarray}
%-----------------------------------------------------------------------------------------------------
where
%-----------------------------------------------------------------------------------------------------
\begin{eqnarray}
N_{9a}(x) &=& \frac{5}{8}(13 + 16 x^2 - x^4) -  \frac{x^2}{2} (24 - x^2) \ln(x)  + 3 x^2 \ln^2(x).
\label{eq:N9a}
\end{eqnarray}
%-----------------------------------------------------------------------------------------------------
For the second system, we obtain
%-----------------------------------------------------------------------------------------------------
\begin{eqnarray}
\left(\frac{d^2}{dx^2}
- \frac{1 - 30 x^2 + 45 x^4}{x(9x^2-1)(x^2-1)} \frac{d}{dx}
- \frac{24 (1 - 3 x^2)}{(9x^2-1)(x^2-1)}\right) f_{8b}(x)
&=& 
N_{8b}(x),
%%\label{eq:two}
\label{eq:f9bc}
\end{eqnarray}
%-----------------------------------------------------------------------------------------------------
where
%-----------------------------------------------------------------------------------------------------
\begin{eqnarray}
N_{8b}(x) &=& 
\frac{1}{\left(9x^2-1\right)\left(x^2-1\right)} \biggl[
-32 x^2 \left(8 - 27 x^2\right) \ln^3(x)
-12 \left(1 - 13 x^2 - 216 x^4 + 162 x^6\right) \ln^2(x)
\nonumber\\ &&
+6 \left(6 - 46 x^2 - 399 x^4 + 81 x^6\right) \ln(x)
+\frac{61}{2} - \frac{415}{2} x^2 + \frac{2199}{2} x^4 + \frac{675}{2} x^6
\biggr]~,
\label{eq:two}
\end{eqnarray}
%-----------------------------------------------------------------------------------------------------
and
%-----------------------------------------------------------------------------------------------------
\begin{eqnarray}
f_{9b}(x) &=& N_{9b}(x)-2 f_{8b}(x)+\frac{x}{2} \frac{d}{dx} f_{8b}(x)~.
\label{eq:f9bb}
\end{eqnarray}
%-----------------------------------------------------------------------------------------------------
with
%-----------------------------------------------------------------------------------------------------
\begin{eqnarray}
N_{9b}(x) &=& 
 9 x^2 \left(1+6 x^2\right) \ln^2(x)
- \frac{9}{2} x^2 \left(8+15 x^2\right) \ln(x) 
+ \frac{15}{8} \left(-1+16 x^2+13 x^4\right)~.
\label{eq:f9b}
\end{eqnarray}
%-----------------------------------------------------------------------------------------------------
In terms of the variable $x^2$, we can see that the differential equations (\ref{eq:one}) 
and (\ref{eq:f9bc}) have four singular points. Three of them are the standard ones of 
hypergeometric functions at $x=0,1,\infty$. The fourth one is at $x^2=9$ in the case of 
(\ref{eq:one}), and at $x^2=1/9$ in the case of (\ref{eq:f9bc}). 

The derivatives of the constant terms $f_{10a}(x)$ and $f_{10b}(x)$ of the master integrals $J^{(3)}_{10a}(x)$ and $J^{(3)}_{10b}(x)$ can be written entirely in terms of harmonic polylogarithms and the constant terms of the previous integrals,
%-----------------------------------------------------------------------------------------------------
\begin{eqnarray}
\label{eq:10aa}
\frac{d}{dx} f_{10a}(x) &=& N_{10a}(x)
+\frac{4}{(x^2-1)^2 x} f_{8a}(x)
+\frac{2 \big(x^2+3\big)}{(x^2-1)^2 x} f_{9a}(x),
\\
%----
\label{eq:10bb}
\frac{d}{dx} f_{10b}(x) &=& N_{10b}(x)
+\frac{4}{3 (x^2-1)^2 x^3} f_{8b}(x)
+\frac{2 \big(
        3 x^2+1\big)}{3 (x^2-1)^2 x^3} f_{9b}(x)~.
\end{eqnarray}
%-----------------------------------------------------------------------------------------------------
The terms $N_{10a}(x)$  and $N_{10b}(x)$ have been given in Ref.~\cite{Ablinger:2017bjx} and contain the harmonic polylogarithms, which are defined by \cite{Remiddi:1999ew}
%-----------------------------------------------------------------------------------------------------
\begin{eqnarray}
\label{eq:HPL}
H_{b,\vec{a}}(x) &=& \int_0^x dy f_b(y) H_{\vec{a}}(y);~~ f_b(x) \in \left\{f_0,f_1, f_{-1}\right\} 
\equiv \left\{\frac{1}{x},\frac{1}{1-x}, \frac{1}{1+x} \right\}; 
\nonumber\\ && H_{\tiny \underbrace{\tiny{0, ...,0}}_k}(x) =\frac{1}{k!} \ln^k(x); \quad H_\emptyset(x) 
\equiv 1~. 
\end{eqnarray}
%-----------------------------------------------------------------------------------------------------

%%%%%%%%%%%%%%%%%%%%%%%%%%%%%%%%%%%%%%%%%%%%%%%%%%%%%%%%%%%%%%%%%%%%%%%%%%%%%%%%%%%%%%%%%%%%%%%%%%%%%%%%%%%%%%%%%%%%%%%%%%%%%%%%%%%%%%
\section{Solutions}

\noindent
The homogeneous solutions of Eq.~(\ref{eq:one}) can be found in terms of hypergeometric 
functions at rational argument using the algorithms presented in 
\cite{IVH,Ablinger:2017bjx}. They read
%-----------------------------------------------------------------------------------------------------
\begin{eqnarray}
\label{eq:ps1a}
\psi_{1a}^{(0)}(x) &=& \sqrt{2 \sqrt{3} \pi} 
\frac{x^2 (x^2-1)^2 (x^2-9)^2}{(x^2+3)^4} \,
\pFq{2}{1}{{\tfrac{4}{3}},\tfrac{5}{3}}{2}{z}
\\
\label{eq:ps2a}
\psi_{2a}^{(0)}(x) &=& \sqrt{2 \sqrt{3} \pi}
\frac{x^2 (x^2-1)^2 (x^2-9)^2}{(x^2+3)^4} \,
\pFq{2}{1}{{\tfrac{4}{3}},\tfrac{5}{3}}{2}{1-z},
\end{eqnarray}
%-----------------------------------------------------------------------------------------------------
with
%-----------------------------------------------------------------------------------------------------
\begin{eqnarray}
z = \frac{x^2(x^2-9)^2}{(x^2+3)^3}~.
\end{eqnarray}
%-----------------------------------------------------------------------------------------------------
The Wronskian for this system is
%-----------------------------------------------------------------------------------------------------
\begin{eqnarray}
\label{eq:W1}
W_a(x) = x (9 - x^2) (x^2-1).
\end{eqnarray}
%-----------------------------------------------------------------------------------------------------
Three of the singularities of the differential equation~(\ref{eq:one}) are encoded by the 
hypergeometric functions, while the remaining singularity shows up in the rational 
argument.

\noindent
Equivalent solutions are found by applying relations due to triangle groups \cite{TAKEUCHI},
%-----------------------------------------------------------------------------------------------------
\begin{eqnarray}
\label{eq:ps1b}
\psi_{1b}^{(0)}(x) &=& \frac{\sqrt{\pi}}{12 \sqrt{2}} \sqrt{(1+x) (3-x)} \Biggl\{
(x-1)(x+3)^2 \,
\pFq{2}{1}{{\tfrac{1}{2}},\tfrac{1}{2}}{1}{z}
\nonumber\\ && 
- (x^2+3)(x-3) \,
\pFq{2}{1}{{\tfrac{1}{2}},-\tfrac{1}{2}}{1}{z} \Biggr\}
\\
\label{eq:ps2b}
\psi_{2b}^{(0)}(x) &=&
\frac{\sqrt{\pi}}{\sqrt{2}} \sqrt{(1+x) (3-x)} \Biggl\{
x^2 \, \pFq{2}{1}{{\tfrac{1}{2}},\tfrac{1}{2}}{1}{1-z}
\nonumber\\ && 
+\frac{1}{8} (x-3)(x^2+3) \,
\pFq{2}{1}{{\tfrac{1}{2}},-\tfrac{1}{2}}{1}{1-z}\Biggr\},
\end{eqnarray}
%-----------------------------------------------------------------------------------------------------
with
%-----------------------------------------------------------------------------------------------------
\begin{equation}
z = -\frac{16 x^3}{(x+1) (x-3)^3}.
\end{equation}
%-----------------------------------------------------------------------------------------------------
These solutions have the same Wronskian as the previous one up to a sign. The hypergeometric functions in Eqs. (\ref{eq:ps1b}) and (\ref{eq:ps2b}) are related to the elliptic integrals of the first and second kind,
%-----------------------------------------------------------------------------------------------------
\begin{eqnarray}
\pFq{2}{1}{{\tfrac{1}{2}},\tfrac{1}{2}}{1}{z}  &=& \frac{2}{\pi} {\bf K}(z), \\
\pFq{2}{1}{{\tfrac{1}{2}},-\tfrac{1}{2}}{1}{z} &=& \frac{2}{\pi} {\bf E}(z)~,
\label{ell:F2EandK}
\end{eqnarray}
%-----------------------------------------------------------------------------------------------------
which have the following integral representations in Legendre's normal form \cite{LEGENDRE},
%-----------------------------------------------------------------------------------------------------
\begin{eqnarray}
{\bf K}(z) &=& \int_0^1 \frac{dt}{\sqrt{(1-t^2)(1-zt^2)}}, \\
{\bf E}(z) &=& \int_0^1 dt \sqrt{\frac{1-zt^2}{1-t^2}}~.
\label{eq:EandKintreps}
\end{eqnarray}
%-----------------------------------------------------------------------------------------------------
In the case of Eq. (\ref{eq:f9bc}), we obtain the following homogeneous solutions
%-----------------------------------------------------------------------------------------------------
\begin{eqnarray}
\label{eq:ps3}
\psi^{(0)}_{3}(x) &=&
-\frac{\sqrt{1-3x} \sqrt{x+1}}{2 \sqrt{2 \pi}} \Biggl[
   (x+1) \left(3 x^2+1\right) {\bf E}\left(z\right)
-  (x-1)^2 (3 x+1) {\bf K}\left(z\right)
\Biggr]
\\
\label{eq:ps4}
\psi^{(0)}_{4}(x) &=&
-\frac{\sqrt{1-3x} \sqrt{x+1}}{2\sqrt{2\pi}} \Biggl[
8 x^2  {\bf K}\left(1-z\right)
- (x+1) \left(3 x^2+1\right) {\bf E}\left(1-z\right)
\Biggr],
\end{eqnarray}
%-----------------------------------------------------------------------------------------------------
where
%-----------------------------------------------------------------------------------------------------
\begin{equation}
z = \frac{16 x^3}{(x+1)^3 (3x-1)},
\end{equation}
%-----------------------------------------------------------------------------------------------------
with the Wronskian
%-----------------------------------------------------------------------------------------------------
\begin{equation}
W_b(x) = x (9 x^2-1) (x^2-1),
\end{equation}
%-----------------------------------------------------------------------------------------------------
The solutions to the inhomogeneous equations can be obtained using the 
method of Euler-Lagrange 
variation of constants. The presence of elliptic integrals in the homogeneous solutions 
(\ref{eq:ps1b}), (\ref{eq:ps2b}), (\ref{eq:ps3}) and (\ref{eq:ps4}) then leads to 
generalized iterated integrals where one of the letters in the alphabet is itself an 
integral that cannot be rewritten in such a way that it becomes also part of 
the iteration chain. These are the so called iterated non-iterated integrals, defined by, 
cf.~\cite{Ablinger:2017bjx}, 
%-----------------------------------------------------------------------------------------------------
\begin{eqnarray}
\label{eq:ITNEW}
\mathbb{H}_{a_1,..., a_{m-1};\{ a_m; F_m(r(y_m))\},a_{m+1},...,a_q}(x) &=& \int_0^x dy_1 
f_{a_1}(y_1) \int_0^{y_1} dy_2 ... \int_0^{y_{m-1}} dy_m f_{a_m}(y_m) F_m[r(y_m)] 
\nonumber\\ &&
\times H_{a_{m+1},...,a_q}(y_{m+1}).
\end{eqnarray}
%------------------------------------------------------------------------------------
One can generalize this even further to cases where more than one definite integral $F_m$ appears. Here the $f_{a_i}(y)$ are the usual letters of the different classes considered in 
\cite{Remiddi:1999ew,Ablinger:2011te,Ablinger:2013cf,Ablinger:2014bra} multiplied by hyperexponential pre-factors 
%------------------------------------------------------------------------------------
\begin{eqnarray}
r(y) y^{r_1}(1-y)^{r_2},~~~~r_i \in \mathbb{Q},~r(y) \in \mathbb{Q}[y]
\end{eqnarray}
%------------------------------------------------------------------------------------
and $F[r(y)]$ is given by
%------------------------------------------------------------------------------------
\begin{eqnarray}
F[r(y)] = \int_0^1 dz g(z,r(y)),~~~r(y) \in \mathbb{Q}[y].
\end{eqnarray}
%------------------------------------------------------------------------------------
We have chosen here $r(y)$ as a rational function because this is what we need for the calculations we are presenting here, but other functions may also be possible. Specifically, we have
%-----------------------------------------------------------------------------------------------------
\begin{eqnarray}
\label{eq:F1}
F[r(y)] &=& \pFq{2}{1}{a,b}{c}{r(y)} = \frac{\Gamma(c)}{\Gamma(b) \Gamma(c-b)} 
\int_0^1 dz z^{b} (1-z)^{c-b-1} \left(1-r(y) z\right)^{-a},
\end{eqnarray}
%------------------------------------------------------------------------------------
with $r(y) \in \mathbb{Q}[y]$, and $a,b,c \in \mathbb{Q}$.

The new iterated integral (\ref{eq:ITNEW}) is not limited to the emergence of the functions (\ref{eq:F1}).
Multiple definite integrals are allowed as well, such as the Appell hypergeometric functions \cite{HYP,SLATER} 
or even more involved higher functions. The integrals defined in (\ref{eq:ITNEW}) also obey relations of the shuffle type~\cite{SHUF,Blumlein:2003gb} with respect to their letters 
$f_{a_m}(y_m) (F_m[r(y_m)])$.

In the case of Eq. (\ref{eq:one}), we get
%-----------------------------------------------------------------------------------------------------
\begin{eqnarray}
f_{8a}(x) &=& \psi^{(0)}_{1b}(x) \biggl\{
C_1 - \int_0^x dz \, \biggl[-\frac{N_{8a}(z)}{W_a(z)} \psi^{(0)}_{2b}(z)
-\sqrt{\frac{3}{2 \pi}} \frac{3}{4 z} \big(43-24 \ln(z)+8 \ln^2(z)\big)\biggr]\biggr\}
\nonumber \\ &&
-\frac{3}{4} \sqrt{\frac{3}{2 \pi}} \psi^{(0)}_{1b}(x) \biggl[43 \ln(x)-12 \ln^2(x)+\frac{8}{3} \ln^3(x)\biggr]
\nonumber \\ &&
+\psi^{(0)}_{2b}(x) \biggl[C_2 + \int_0^x dz \, \biggl(-\frac{N_{8a}(z)}{W_a(z)} \psi^{(0)}_{1b}(z)\biggr)\biggr]~,
\label{eq:f8a}
\end{eqnarray} 
%-----------------------------------------------------------------------------------------------------
with the integration constants
%-----------------------------------------------------------------------------------------------------
\begin{eqnarray}
C_1 &=& 
-18 \sqrt{\frac{2}{\pi }} {\sf Im}\Biggl[\text{Li}_3\left(\frac{e^{-\frac{i \pi}{6}}}{\sqrt{3}}\right)\Biggr]
-\frac{35 \pi ^{5/2}}{36 \sqrt{2}}
-\frac{25}{8} \sqrt{\frac{3}{2 \pi }}
-\frac{3}{4} \sqrt{\frac{\pi }{2}} \ln^2(3)
\nonumber \\ &&
-2 \sqrt{\frac{2}{3}} \pi ^{3/2} \ln(3)
+\frac{45}{4} \sqrt{\frac{3}{2 \pi }} \ln(3)
+\sqrt{\frac{6}{\pi }} \ln(3) \psi'\left(\frac{1}{3}\right),
\\
C_2 &=&
\frac{1}{36} \sqrt{\frac{\pi}{6}} \left[-135+16 \pi^2-24 
\psi'\left(\frac{1}{3}\right)\right].
\end{eqnarray} 
%-----------------------------------------------------------------------------------------------------
The second term in the first integral in Eq. (\ref{eq:f8a}) had to be introduced in order to regulate the singularity at $z=0$ of the integrand. The corresponding result for this term as an indefinite integral is then subtracted in the second line of (\ref{eq:f8a}), accordingly.

We can now obtain the solutions for $f_{9a}(x)$  by inserting the result (\ref{eq:f8a})
 in (\ref{eq:f9aa}).
%-----------------------------------------------------------------------------------------------------
\begin{eqnarray}
f_{9a}(x) &=& P_1(x) \biggl\{
C_1 - \int_0^x dz \, \biggl[-\frac{N_{8a}(z)}{W_a(z)} \psi^{(0)}_{2b}(z)
-\sqrt{\frac{3}{2 \pi}} \frac{3}{4 z} \big(43-24 \ln(z)+8 \ln^2(z)\big)\biggr]\biggr\}
\nonumber \\ &&
-\frac{3}{4} \sqrt{\frac{3}{2 \pi}} P_1(x) \biggl[43 \ln(x)-12 \ln^2(x)+\frac{8}{3} \ln^3(x)\biggr]
\nonumber \\ &&
+P_2(x) \biggl[C_2 + \int_0^x dz \, \biggl(-\frac{N_{8a}(z)}{W_a(z)} \psi^{(0)}_{1b}(z)\biggr)\biggr]
+N_{9a}(x),
\end{eqnarray} 
%-----------------------------------------------------------------------------------------------------
where
%-----------------------------------------------------------------------------------------------------
\begin{eqnarray}
P_i(x) &=& \frac{x}{6} \frac{d}{dx} \psi^{(0)}_{ib}(x) - \frac{2}{3} \psi^{(0)}_{ib}(x), \quad i=1,2.
\end{eqnarray} 
%-----------------------------------------------------------------------------------------------------
Inserting these solutions in Eqs. (\ref{eq:10aa}) and integrating 
in 
$x$, we obtain the solution for $f_{10a}(x)$. 
%-----------------------------------------------------------------------------------------------------
\begin{eqnarray}
f_{10a}(x) &=& \int_0^x dz \biggl[N_{10a}(z)+\frac{2 (3+z^2)}{z (1-z^2)^2} N_{9a}(z)\biggr]
\nonumber \\ &&
+\int_0^x dz \frac{4}{z (1-z^2)^2} \Biggl\{
\biggl(\psi^{(0)}_{1b}(z)+\frac{1}{2} (3+z^2) P_1(z)\biggr) \biggl[C_1
\nonumber \\ &&
-\frac{3}{4} \sqrt{\frac{3}{2 \pi}} \biggl(43 \ln(z)-12 \ln^2(z)+\frac{8}{3} \ln^3(z)\biggr)\biggr]
+C_2 \biggl(\psi^{(0)}_{2b}(z)+\frac{1}{2} (3+z^2) P_2(z)\biggr)
\Biggr\}
\nonumber \\ &&
+\int_0^x dy \int_0^y dz \, \frac{4}{y (1-y^2)^2} \Biggl\{
\frac{N_{8a}(z)}{W_a(z)} \biggl[
\psi^{(0)}_{1b}(y) \psi^{(0)}_{2b}(z) - \psi^{(0)}_{1b}(z) \psi^{(0)}_{2b}(y)
\nonumber \\ &&
-\frac{1}{2} (3+y^2) \big(P_1(y) \psi^{(0)}_{2b}(z)+P_2(y) \psi^{(0)}_{1b}(z)\big)\biggr] 
\nonumber \\ &&
+\sqrt{\frac{3}{2 \pi}} \biggl[\psi^{(0)}_{1b}(y)-\frac{1}{2} (3+y^2) P_1(y)\biggr] \frac{3}{z} \biggl(43 \ln(z)-12 \ln^2(z)+\frac{8}{3} \ln^3(z)\biggr)
\Biggr\}
\nonumber \\ &&
-\frac{19}{72} \pi^4+\frac{2}{3} \pi^2 \psi'\left(\frac{1}{3}\right)-\frac{1}{2} \psi'\left(\frac{1}{3}\right)^2+6 \zeta(3).
\label{eq:f10a}
\end{eqnarray} 
%-----------------------------------------------------------------------------------------------------
Some of the terms appearing in (\ref{eq:f10a}) have been introduced 
(and appropriately subtracted) in order to regulate the singularities of the integrands, 
like we did in the case of Eq. (\ref{eq:f8a}). The last line of (\ref{eq:f10a}) 
corresponds to the integration constant of Eq. (\ref{eq:10aa}), obtained from boundary conditions. Notice that now we have a double integral over the homogeneous elliptic solutions, which is still numerically stable when evaluated for a specific numerical value of $x$. All of our solutions can be expanded around $x=0$ and $x=1$, and they agree with the expansions given in \cite{Grigo:2012ji}.
In a similar way, one obtains the solutions of $f_{8b, 9b, 10b}$, 
cf.~\cite{Ablinger:2017bjx}.

We inserted our solutions in the expression for the $\delta^{(2)}$, see 
Eq.~(\ref{eq:deltarho}),
in terms of the master integrals in the $\overline{\rm MS}$ scheme. The contribution 
with respect to the iterative non-iterative integrals is given by 
\begin{eqnarray}
\delta^{(2)}(x) &=& \dots + C_F \left(C_F-\frac{C_A}{2}\right) \Biggl[
\frac{11-x^2}{12 (1-x^2)^2} f_{8a}(x)
+\frac{9-x^2}{3 (1-x^2)^2} f_{9a}(x)
+\frac{1}{12} f_{10a}(x)
\\ && 
+\frac{5-39 x^2}{36 (1-x^2)^2} f_{8b}(x)
+\frac{1-9 x^2}{9 (1-x^2)^2} f_{9b}(x)
+\frac{x^2}{12} f_{10b}(x)
\Biggr]
\\ &&
+\frac{C_F T_F}{9 (1-x^2)^3} \Biggl[
(5 x^4-28 x^2-9) f_{8a}(x)
+\frac{1-3 x^2}{3 x^2} (9 x^4+9 x^2-2) f_{8b}(x)
\\ && 
+(9-x^2) (x^4-6 x^2-3) f_{9a}(x)
+\frac{1-9 x^2}{3 x^2} (3 x^4+6 x^2-1) f_{9b}(x)
\Biggr].
\end{eqnarray}
The color factor signals that it stems from the non-planar part of the problem.
%-----------------------------------------------------------------------------------------------------
\begin{figure}[t]
\begin{center}
\includegraphics[width=0.7\textwidth]{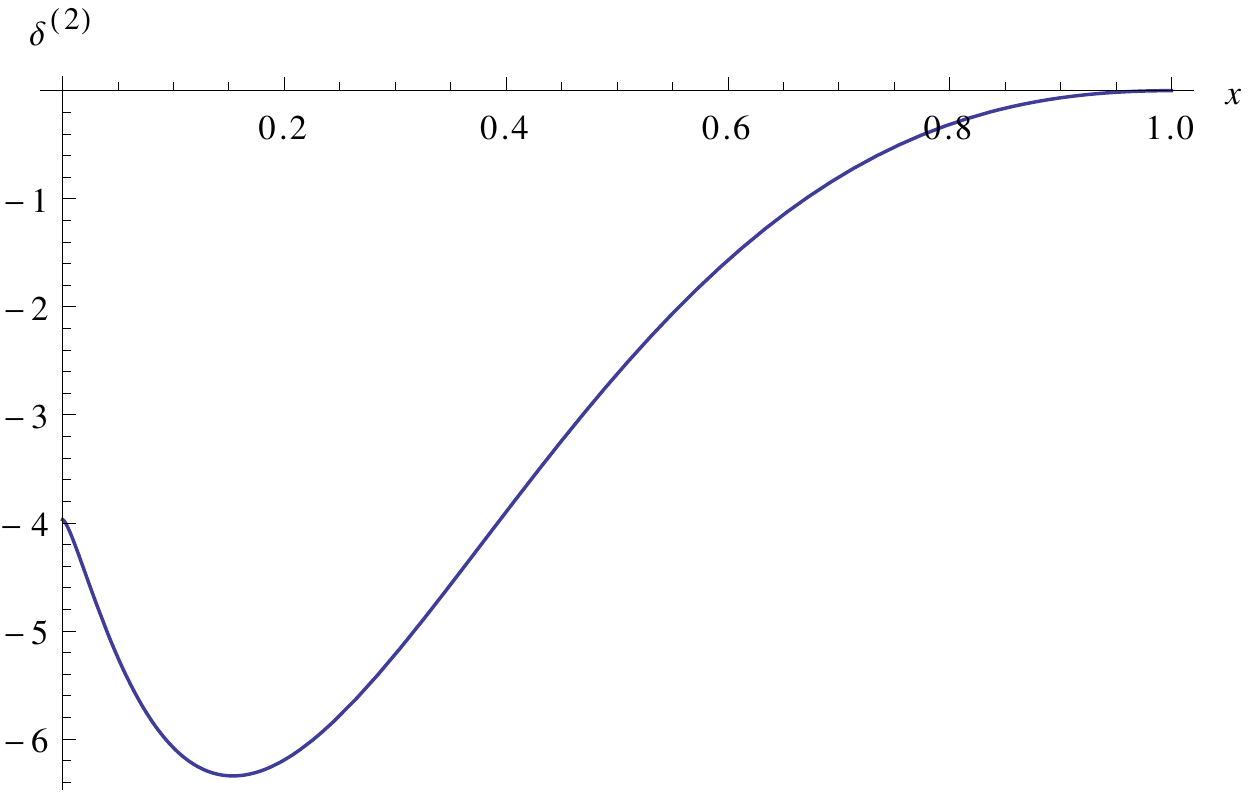}
\end{center}
\caption{The two-mass contributions to $\delta^{(2)}$ as a function of $x$.}
\label{fig:delta2}
\end{figure}
%-----------------------------------------------------------------------------------------------------

In Figure~\ref{fig:delta2}, 
we can see the resulting plot as a function of $x$. We see that $\delta^{(2)}(x) 
\rightarrow 0$ as $x \rightarrow 1$, as expected, and for $x=0$, we obtain $\delta^{(2)}(0) 
= -3.9696$, which agrees with the result presented in \cite{Chetyrkin:1995ix} in the limit 
of a small mass ratio.

%%%%%%%%%%%%%%%%%%%%%%%%%%%%%%%%%%%%%%%%%%%%%%%%%%%%%%%%%%%%%%%%%%%%%%%%%%%%%%%%%%%%%%%%%%%%%%%%%%%%%%%%%%%%%%%%%%%%%%%%%%%%%%%%%%%%%%%%%%%%%%%%%%%%%%%%%%
\section{\boldmath $\eta$-ratios and $\mathbf{q}$-series representations}

\noindent
The appearance of complete elliptic integrals calls for the study of their representation 
in terms of 
related functions such as the Dedekind $\eta$ function, 

\begin{equation}
\eta(\tau) = q^{\frac{1}{12}} \prod_{k=1}^{\infty} \left(1-q^{2 k}\right),
\end{equation}
and the Jacobi $\vartheta_i$ functions

\begin{equation}
\vartheta_2(q) = \frac{2 \eta^2(2\tau)}{\eta(\tau)}, \quad 
\vartheta_3(q) = \frac{\eta^5(\tau)}{\eta^2\left(\frac{1}{2}\tau\right) \eta^2(2 \tau)}, \quad
\vartheta_4(q) = \frac{\eta^2\left(\frac{1}{2}\tau\right)}{\eta(\tau)},
\end{equation}

\noindent
with $q=\exp(i \pi \tau)$ and $\tau$ located in the complex upper half-plane.
Applying a higher order Legendre-Jacobi transformation \cite{BB,Broadhurst:2008mx},
one may transform the variable $x$ in ${\bf K}(k^2) \equiv {\bf K}(r(x))$ into the nome $q$ analytically by

\begin{equation}
k^2=r(x)=\frac{\vartheta_2^4(q)}{\vartheta_3^4(q)}. 
\end{equation}
The integrands can then be written in terms of products of meromorphic modular forms, which 
can be rewritten
in terms of linear combinations of ratios of $\eta$ functions.

One may express the elliptic integral of the first kind ${\bf K}$ appearing in the homogeneous solutions by

\begin{equation}
{\bf K}(k^2) = \frac{\pi \eta^{10}(\tau)}{2 \eta^4\left(\frac{\tau}{2}\right) \eta^4(2\tau)}
\end{equation}
and
\begin{equation}
{\bf E}(k^2) = {\bf K}(k^2)+\frac{\pi^2 q}{{\bf K}(k^2)} \frac{d}{dq} \ln\left(\vartheta_4(q)\right).
\end{equation}

Other terms appearing, e.g., in $\psi^{(0)}_3(x)$ and $\psi^{(0)}_4(x)$, such as $\sqrt{(1-3 x) (1+x)}$ can also be expressed in terms 
of $\eta$ ratios:

\begin{equation}
\sqrt{(1-3x)(1+x)} = \frac{i}{\sqrt{3}} \left. 
\frac{
\eta\left(\tfrac{\tau}{2}\right)
\eta\left(\tfrac{3\tau}{2}\right) \eta(2\tau) \eta(3\tau)} 
{\eta(\tau)\eta^3(6\tau)}\right|_{q\rightarrow -q}.
\end{equation}
All other ingredients in the homogeneous solutions can be expressed in similar ways.
In the case of one of the homogeneous solutions to the differential equation of $f_{8a}(x)$, namely

\begin{equation}
\psi^{(0)}_{1b}(x) = \frac{2}{\sqrt{3}} H(x)
\end{equation}
with
\begin{equation}
H(x) = \frac{x^2 (x^2-1)^2 (x^2-9)^2}{(x^2+3)^4} \pFq{2}{1}{\frac{4}{3},\frac{5}{3}}{2}{\frac{x^2 (x^2-9)^2}{(x^2+3)^3}}
\end{equation}
by setting the kinematic variable
\begin{equation}
x = 3 \frac{\eta_1^2 \eta_6^4}{\eta_2^4 \eta_3^2}, 
\end{equation}
Broadhurst \cite{Bro} has found the following modular representation:

\begin{eqnarray}
H\left(3 \frac{\eta_1^2 \eta_6^4}{\eta_2^4 \eta_3^2}\right) &=&
\frac{1}{2} \left[
\frac{\eta_1^{14} \eta_6^{10}}{\eta_2^{22} \eta_3^2}
+\frac{\eta_1^6 \eta_6^4}{\eta_2^{12} \eta_3^2} \left(
\frac{\eta_1^4 \eta_6^8}{\eta_2^8 \eta_3^4}+\frac{1}{3}
\right) q \frac{d}{dq}\right] \frac{\eta_2 \eta_3}{\eta_1^3 \eta_6^2} 
\label{eq:Bro}
\\ &=&
q-6 q^2+24 q^3-74 q^4+195 q^5-474 q^6+1100 q^7+{\cal O}(q^8), \nonumber
\end{eqnarray}
where $\eta_k = \eta(k \tau)$.

The inhomogeneities can be dealt with in a similar way. For example, in the case of $f_{8b}(x)$, the inhomogeneous solution is of the form

\begin{equation}
I = \sum_{m=1}^8 c_m \int \frac{dx}{x} H_0^n(x) \hat{f}_m(x) \psi^{(0)}_{3,4}(x),~~~n 
\in \{0,1,2,3\},~~
c_m \in \mathbb{Q},
\end{equation}
with

\begin{equation}
\hat{f}_m \in \left\{
\frac{1}{1 \pm x}, 
\frac{1}{(1 \pm x)^2}, 
\frac{1}{1 \pm 3x},
\frac{1}{(1 \pm 3x)^2}\right\}.
\end{equation}
One obtains the following $\eta$ ratios

\begin{eqnarray*}
\frac{1}{1-x} &=& -3 \frac{\eta^2(\tau) \eta\left(\tfrac{3}{2}\tau\right) 
\eta^3(6\tau)}{\eta^3\left(\tfrac{1}{2}\tau\right) \eta(2\tau) \eta^2(3\tau)}
\\ && 
\\
\frac{1}{1-3 x} &=& -\frac{\left[\eta(\tau) \eta\left(\tfrac{3}{2}\tau\right) 
\eta^2(6\tau)\right]^3}{\eta\left(\tfrac{1}{2}\tau\right) \eta^2(2\tau) \eta^9(3\tau)}.
\end{eqnarray*}
Writing the solutions to our differential equations in terms of this type of functions 
has the advantage that all singular points are treated in the same way, 
unlike the expressions presented in the previous section, where one of the singularities 
is treated differently and appears as a singularity of the rational function in the 
argument of the elliptic integrals. 

In conclusion, we mention that the concept of iterative integrals in solving 
Feynman-parameter integrals analytically finds its generalization in the so-called
iterative non-iterative integrals. Here new letters emerge, depending on the next 
integration variables, which are given by (multiple) integral representations, which 
cannot be rewritten in terms of iteratative integrals themselves.

\vspace*{3mm}
\noindent
{\bf Acknowledgment.}~~We would like to thank J.~Ablinger, D.~Broahurst, H.~Cohen, 
C.~Raab,  C.-S.~Radu, and D.~Zagier for 
discussions. This work has been support in part Austrian Science Fund (FWF) grant SFB F50 
(F5009-N15).

%%%%%%%%%%%%%%%%%%%%%%%%%%%%%%%%%%%%%%%%%%%%%%%%%%%%%%%%%%%%%%%%%%%%%%%%%%%%%%%%%%%%%%%%%%%%%%%%%%%%%%%%%%%%%%%%%%%%%%%%%%%%%%%%%%%%%%%%%%%%%%%%%%%%%%%%%%

%-----------------------------------------------------------------------------------------------------
%-----------------------------------------------------------------------------------------------------
\end{document}